\documentclass[conference]{IEEEtran}
\IEEEoverridecommandlockouts
\usepackage{bbm}
\usepackage{amsfonts}
\usepackage[dvips]{graphicx}
\usepackage{times}
\usepackage{cite}
\usepackage{amsmath}
\usepackage{array}
\usepackage{amssymb}

\newcommand{\bs}{\boldsymbol}

\usepackage{stfloats}
\usepackage{graphicx}
\usepackage{footnote}
\usepackage{booktabs}
\usepackage{array}
\usepackage{algorithm}
\usepackage{subeqnarray}
\usepackage{cases}
\usepackage{threeparttable}
\usepackage{color}
\usepackage{hyperref}
\usepackage{epstopdf}
\usepackage{algpseudocode}
\usepackage{bm}
\usepackage{multirow}
\usepackage[labelformat=simple]{subcaption}
\usepackage[font={small}]{caption}
\usepackage{adjustbox}

\usepackage{enumitem}

\makeatletter
\newcommand\fs@betterruled{%
  \def\@fs@cfont{\bfseries}\let\@fs@capt\floatc@ruled
  \def\@fs@pre{\vspace*{5pt}\hrule height.8pt depth0pt \kern2pt}%
  \def\@fs@post{\kern2pt\hrule\relax}%
  \def\@fs@mid{\kern2pt\hrule\kern2pt}%
  \let\@fs@iftopcapt\iftrue}
\floatstyle{betterruled}
\restylefloat{algorithm}
\makeatother

 \def\BibTeX{{\rm B\kern-.05em{\sc i\kern-.025em b}\kern-.08em T\kern-.1667em\lower.7ex\hbox{E}\kern-.125emX}}

\newlength{\bottomMargin}
\newtheorem{theorem}{\bf Theorem}
\newtheorem{remark}{\bf Remark}

\newtheorem{lemma}{\bf Lemma}
\newtheorem{definition}{\bf Definition}

\allowdisplaybreaks

\begin{document}

\title{Risk-Aware Optimization of Age of Information in the Internet of Things}


\author{\IEEEauthorblockN{Bo Zhou$^{*}$, Walid Saad$^{*}$,  Mehdi Bennis$^{\dagger}$, and Petar Popovski$^{\ddagger}$}
\IEEEauthorblockA{$^{*}$Wireless@VT, Bradley Department of Electrical and Computer Engineering, Virginia Tech, Blacksburg, VA, USA,\\
$^{\dagger}$Center for Wireless Communications, University of Oulu, Finland,\\
$^{\ddagger}$Department of Electronic Systems, Aalborg University, Aalborg, Denmark,\\
Emails: $^{*}$\{ecebo, walids\}@vt.edu, $^{\dagger}$mehdi.bennis@oulu.fi, $^{\ddagger}$petarp@es.aau.dk.
}
\thanks{This research was supported  by  the U.S. National Science Foundation under Grant CNS-1836802,  the Academy of Finland Project CARMA, the Academy of Finland Project MISSION, the Academy of Finland Project SMARTER, the INFOTECH Project NOOR, the European Research Council (ERC) under the European Union Horizon 2020 research and innovation program (ERC Consolidator Grant Nr. 648382 WILLOW), and the Danish Council for Independent Research (Grant Nr. 8022-00284B SEMIOTIC).}
\\[-5.0ex]
}

\maketitle

\begin{abstract}


Minimization of the expected value of age of information (AoI) is a \emph{risk-neutral} approach, and it thus cannot capture rare, yet critical, events with potentially large AoI. In order to capture the effect of these events, in this paper, the notion of conditional value-at-risk (CVaR) is proposed as an effective coherent risk measure that is suitable for minimization of AoI for real-time IoT status updates. In the considered monitoring system, an IoT device monitors a physical process and sends the
status updates to a remote receiver with an updating cost. The optimal status update process is designed to jointly minimize the AoI at the receiver, the CVaR of the AoI at the receiver, and the energy cost. This stochastic optimization problem is formulated as an infinite horizon discounted risk-aware Markov decision process (MDP), which is computationally intractable due to the time inconsistency of the CVaR. By exploiting the special properties of coherent risk measures, the risk-aware MDP is reduced to a standard MDP with an augmented state space, for which we derive the optimal stationary policy using dynamic programming. In particular, the optimal history-dependent policy of the risk-aware MDP is shown to depend on the history only through the augmented system states and can be readily constructed using the optimal stationary policy of the augmented MDP. The proposed solution is shown to be computationally tractable and able to minimize the AoI in real-time IoT monitoring systems in a risk-aware manner.

\end{abstract}

\section{Introduction}
Time-sensitive  Internet  of  Things  (IoT)  applications \cite{8869705}, such as real-time surveillance and monitoring, drone navigation, and autonomous driving, must rely on a timely delivery of status information updates of the physical processes that are being monitored or operated by the IoT devices for control and monitoring purposes.
In light of this, the concept of \emph{age of information} (AoI) has been recently proposed to evaluate the freshness of the status updates at the information destination (e.g., an IoT control center or base station)\cite{6195689,8000687}. 
The AoI is a performance metric that quantifies the  time elapsed since the latest received status update at the information destination was generated. 
Since the AoI captures the  information freshness from the perspective of the remote destination and depends on both the generation and transmission of the status updates, it is fundamentally different from conventional performance metrics, such as throughput or delay. 

Recently, there has been a  growing body of research on minimizing the AoI in various communication systems\cite{8778671,wang2018preempt,8938128,8845182,feng2018age,abd2019reinforcement}.
In \cite{8778671}, the authors study the optimal status sampling and updating policy to minimize the average AoI for an IoT monitoring system under device energy constraints.
The problem of AoI minimization for IoT monitoring systems with non-uniform status packet sizes is studied in \cite{wang2018preempt} and \cite{8938128}.
The works in \cite{8845182,feng2018age} investigate the problem  of AoI minimization for wireless status updating systems with noisy channels.
The authors in \cite{abd2019reinforcement} propose an online sampling policy to minimize the average AoI for energy harvesting systems.

These existing works, e.g.,\cite{8778671,wang2018preempt,8938128,8845182,feng2018age,abd2019reinforcement}, adopt a \emph{risk-neutral} approach, by focusing only on minimizing the expected value of the (random) AoI cost functions, e.g., the average AoI, the average peak AoI, and the average age penalty. 
Although the obtained algorithms through this approach can result in AoI performance that is minimized over a long run, they do not capture the risk of the uncertainty of the AoI cost function, e.g., the variability of the AoI distribution and the effects of rare but potentially detrimental AoI events. 
For example, for safety and state monitoring in industrial production scenarios, a certain status update with a very large AoI could result in a complete shutdown of the production. Thus, it is critical to  focus on the AoI not only in the average sense, but also in a risk-related sense.
Recently, the works in \cite{8640078} and \cite{8937801} considered the tail of the AoI distribution (with extremely large AoI) for vehicular networks and wireless industrial networks, respectively. 
Meanwhile, the work in \cite{8836117} analyzed the violation probability of the peak AoI for a point-to-point communication system with short packets.
However, these approaches in \cite{8640078,8937801,8836117} focus on the probability that the peak AoI exceeds a certain threshold, and, thus, they can neither quantify nor minimize the expected losses that 
might be incurred in tail events in which the AoI is very large. 
Clearly, how to design the optimal status updating policy so as to jointly minimize the average AoI and the expected tail loss of the AoI, remains an open problem.

The main contribution of this paper is a novel design of a risk-aware status updating control policy that jointly minimizes the AoI at the receiver, the expected tail loss of the AoI at the receiver, and the energy cost, for a real-time IoT monitoring system.
Specifically, we use a popular and effective coherent risk measure, called the conditional value-at-risk (CVaR)\cite{rockafellar2000optimization}, to measure the \emph{tail average} of the AoI distribution exceeding a given risk level.
We formulate this stochastic control problem as an infinite horizon discounted risk-aware Markov decision process (MDP) and seek the optimal history-dependent updating control policy. 
This risk-aware MDP is challenging to solve due to two reasons: 1) Because of the time inconsistency of the CVaR, dynamic programming cannot be directly applied and 2) History-dependent policies are generally intractable due to the substantial requirements on the computation time and memory.
By exploiting the dual representation and the temporal decomposition properties of the coherent risk measures, we reduce the risk-aware MDP to a standard MDP on the state space augmented by a two-dimensional risk level space and propose a dynamic programming based solution to derive the \emph{optimal stationary policy} through a risk-aware Bellman operator.
Thus, instead of working on the intractable space of history-dependent policies, it is sufficient to focus on the optimization over stationary policies of the augmented MDP.
In particular, we show that the optimal history-dependent policy depends on the history only through the dynamics of the two risk levels, and can be constructed with the optimal stationary policy for the augmented MDP.
The proposed solution can explicitly account for rare events with very large AoI in an IoT monitoring system and is computationally tractable to obtain the generally intractable history-dependent policies. 

\begin{figure}[!t]
\begin{centering}
\includegraphics[scale=.32]{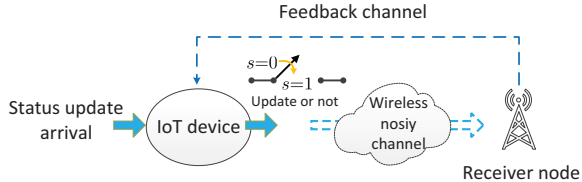}
 \caption{Illustration of a real-time monitoring system.}\label{fig:system}
\end{centering}
\vspace{-0.5cm}
\end{figure}

\section{System Model} 
We consider a general real-time IoT monitoring system composed of an IoT device and a remote receiver node (see Fig.~\ref{fig:system}).
The IoT device  monitors an underlying time-varying physical process and sends the associated real-time status information to the receiver.
We assume that the status information updates of the underlying  process arrive at the IoT device stochastically and are queued at the device before being transmission to the receiver.
We consider a discrete-time system with time slots indexed by $t=0,1,2,\cdots$. 
At the beginning of each time slot, the status update (if any) of the underlying process arrives at the IoT device randomly. 
Similar to \cite{wang2018preempt} and \cite{8938128}, the process of the status update arrivals is modeled by an independent and identically distributed (i.i.d.)  Bernoulli process with mean rate $\lambda\in[0,1]$.
The device is equipped with a buffer to store the arriving status update and the newly arriving most up-to-date status update will replace the older one (if any) in the buffer, as the receiver will not benefit from obtaining an outdated status update. Hence, there is at most one status update at the device.

We consider a wireless packet erasure channel between the IoT device and the receiver, and, upon transmission, each status update will be successfully delivered to the receiver with probability $p$.
 As in \cite{8845182,feng2018age,8938128}, we further assume that the IoT device will be notified immediately upon a successful transmission, through a perfect feedback channel between the device and the receiver.

\subsection{Monitoring Model}
Due to the possible failure of each transmission, the status update currently in the buffer at the device may be outdated at the receiver. 
Thus, in each slot, the IoT device must decide whether to transmit the locally available status update or stay idle to wait for a possibly arriving fresher status update. 
Let $s_t\in\mathcal{S}\triangleq\{0,1\}$ be the updating control action of the device at slot $t$, where $s_t=1$ implies that the device transmits its locally available status update at slot $t$ and $s_t=0$ indicates the device stays idle. $\mathcal{S}$ denotes the control action space.
Let $C$ be the energy cost for transmitting a status update.

\subsection{Age of Information Model}
We adopt the AoI as the key performance metric to quantify the freshness of the status information update at the receiver. The AoI is defined as time elapsed since the most recent status update delivered at the receiver. 
Let $A_{r,t}$ be the AoI at the receiver at the beginning of time slot $t$. By definition, we have $A_{r,t} = t - U^r_t$, where $U^r_t$ is the time stamp of the freshest status update that was delivered to the receiver before $t$.
Note that, the device can only transmit its currently available status update to the receiver and, thus, the AoI at the receiver depends on the age of the status update in the buffer at the device. 
We define $A_{d,t}$ as the AoI at the device at the beginning of slot $t$, to capture the freshness of the status information update at the device. 
Let $\hat{A}_d$ and $\hat{A}_r$ be the upper limits of the AoI at the device and the AoI at the receiver, respectively.
Since  a status update with an infinite age is not meaningful for real-time IoT monitoring systems, we  assume that  $\hat{A}_d$ and $\hat{A}_r$ are finite. Mathematically, $\hat{A}_d$ and $\hat{A}_r$ can be arbitrarily large.
Let $\mathcal{A}_d\triangleq \{1,2,\cdots,\hat{A}_d\}$  and $\mathcal{A}_r\triangleq \{1,2,\cdots,\hat{A}_r\}$ be, respectively, the state space of the AoI at the device and the AoI at the receiver. 
We denote by $\bs{A}_t\triangleq (A_{d,t},A_{r,t})\in\mathcal{A}\triangleq \mathcal{A}_d\times\mathcal{A}_r$ the system AoI state at slot $t$, where $\mathcal{A}$ is the system AoI state space.

Now, we present how $\bs{A}_t$ evolves with the updating control action $s_t$. 
For the AoI at the device, if there is a status update arriving at the device during slot $t$, the AoI at the device will be reset to one, otherwise, the AoI will increase by one. 
Then, the dynamics of $A_{d,t}$ will be given by:
\par\nobreak 
\vspace{-10pt}
{\small
\begin{align}\label{eqn:aoi_device}
A_{d,t+1}=\begin{cases} 1,  &\text{if an update arrives at $t$},\\
                \min\{A_{d,t}+1,\hat{A}_d\}, &\text{otherwise.}
      \end{cases}
\end{align}
}For the AoI at the receiver, if the device transmits the status update to the receiver at slot $t$ and the transmission is successful, then the AoI at the receiver in the next slot will be the current AoI at the device plus one (due to the one slot transmission), otherwise, the AoI will increase by one. Note that, the latter case includes the scenarios in which, the device attempts to send the status update while fails, or the device decides to stay idle. Thus, we have the dynamics of $A_{r,t}$:
\par\nobreak 
\vspace{-10pt}
{\small
\begin{align}\label{eqn:aoi_receiver}
A_{r,t+1}
=\begin{cases} &\min\{A_{d,t}+1,\hat{A}_r\},  ~\text{if}~s_t=1~\text{and the update}\\
&\hspace{25mm}\text{transmission succeeds at}~t,\\
                &\min\{A_{r,t}+1,\hat{A}_r\}, ~ \text{otherwise.}
      \end{cases}
\end{align}
}By comparing \eqref{eqn:aoi_device} and \eqref{eqn:aoi_receiver}, we observe that $A_{r,t}\geq A_{d,t}$ holds for all $t$, and, hence, we only need to focus on the system AoI state space $\mathcal{A}$ with $A_r\geq A_d$. Moreover, we also see that if $A_{d,t}=A_{r,t}$ for some $t$, then there is no need to choose the action $s_t=1$, as the currently available status update at the device has already been delivered to the receiver before $t$.

\section{Problem Formulation}
The existing literature, e.g., \cite{abd2019reinforcement,8778671,8845182,feng2018age,8938128,wang2018preempt}, focuses only on minimizing the expected value of the (random) AoI cost function, and, thus, fails to capture the variability of the AoI distribution and accounts for rare events with potentially very large AoI.
Hence, we consider a \emph{risk-aware} approach, by taking into account the expected value of the AoI and the expected tail loss of the AoI based on the CVaR\cite{rockafellar2000optimization} -- a popular and effective risk measure.

\subsection{Preliminaries on CVaR and risk measures}
For a bounded-mean random variable $Z$ on a probability space $(\Omega, \mathcal{F}, P)$, the CVaR of $Z$ at risk level $\alpha\in(0,1]$ is defined as the expectation of $Z$ in  its $\alpha$-tail distribution\cite{rockafellar2000optimization}:
\begin{align}
\text{CVaR}_{\alpha}(Z) \triangleq \min_{q\in\mathbb{R}}\left\{q+\frac{1}{\alpha}\mathbb{E}[\max\{Z-q,0\}]\right\},
\end{align}
where the expectation is taken over the probability distribution $P$.
Note that, $\text{CVaR}_{\alpha}(Z)$ decreases with $\alpha$, $\text{CVaR}_{1}(Z)= \mathbb{E}[Z]$, and $\lim_{\alpha\to 0}\text{CVaR}_{\alpha}(Z) = \sup(Z)$. Thus, $\alpha$ can be seen as a kind of degree of risk aversion.
It has been shown that the CVaR is a \emph{coherent} risk measure\cite{doi:10.1111/1467-9965.00068,doi:10.1137/1.9780898718751,doi:10.1287/moor.2015.0747}. 
\begin{definition}\label{definition:coherent risk measures}
A \emph{coherent risk measure} $\rho(Z)$ is a mapping from the space $\mathcal{Z}$ of the random variable $Z$ to $\mathbb{R}$ that obeys the following four axioms\cite{doi:10.1111/1467-9965.00068,doi:10.1137/1.9780898718751,doi:10.1287/moor.2015.0747}. For any $Z,Z'\in\mathcal{Z}$:
\begin{enumerate}[label=\arabic*)]
\item Monotonicity: if  $Z\leq Z'$, then $\rho(Z)\leq \rho(Z')$;
\item Subadditivity: $\rho(Z+Z)\leq \rho(Z) + \rho(Z')$;
\item Translation invariance:  $\rho(Z + a) = \rho (Z) + a, \forall a\in\mathbb{R}$;
\item Positive homogeneity: if $b>0$, then $\rho(bZ)=b\rho(Z)$.
\end{enumerate}
\end{definition}

Note that, based on the translation invariance and positive homogeneity axioms of a coherent risk measure $\rho$, we can easily obtain $\rho(c) = c$ for all constants c. One important result in risk measure theory is that each coherent risk measure has its dual representation as the maximum of certain expected value over a risk envelope\cite[Theorem 6.4]{doi:10.1137/1.9780898718751}, i.e.,
\begin{equation}
\rho(Z)  = \max_{\xi\in\Xi} \mathbb{E}_{\xi}[Z],\label{eqn:dual_rep_risk}
\end{equation}
where $\mathbb{E}_{\xi}[Z]\triangleq\sum_{\omega\in\Omega}\xi(\omega)P(\omega)Z(\omega)$ denotes the $\xi$-weighed expectation of $Z$ and $\Xi$ is a specific set of probability density functions, referred to as the risk envelop.
For example, the  risk envelop of the CVaR is $\Xi = \{\xi:\xi(\omega)\in[0,1/\alpha], \forall \omega\in\Omega,~\text{and}~\sum_{\omega\in\Omega}\xi(\omega)P(\omega)=1\}$.

\subsection{Risk-Aware MDP Formulation}
We consider history-dependent updating control policies, and the updating action at each time slot depends on the past history of the system, as represented by the sequence of the previous system AoI states and updating actions.
For each time slot $t=0,1,\cdots$, let $\bs{h}_t\triangleq(\bs{A}_0,s_0,\bs{A}_1,s_1,\cdots,\bs{A}_{t-1},s_{t-1},\bs{A}_t)\in\mathcal{H}_t$ be the history up to  slot $t$, which satisfies the recursion $\bs{h}_t=(h_{t-1},s_{t-1},\bs{A}_t)$ for all $t\geq 1$.
Here, $\mathcal{H}_t$ is the space of all histories up to slot $t$, where $\mathcal{H}_0\triangleq\mathcal{A}$ and $\mathcal{H}_t \triangleq \mathcal{H}_{t-1}\times\mathcal{S}\times\mathcal{A}$ for all $t\geq 1$.
\begin{definition}\label{definition:history-dependent policy}
A \emph{history-dependent updating control policy} $\pi$ is a sequence of decision rules for each time slot, i.e., $\pi\triangleq(\mu_0,\mu_1,\cdots)$, where $\mu_t$  is a mapping from the set of histories $\mathcal{H}_t$ at slot $t$ to the control action space $\mathcal{S}$, i.e., $s_t=\mu_t(\bs{h}_t)$. Let $\Pi_H$ be the set of all history-dependent policies $\pi$.
\end{definition}

By the dynamics in \eqref{eqn:aoi_device} and \eqref{eqn:aoi_receiver}, and the i.i.d. assumptions on the status updates arrival process, the induced random process $\{\bs{A}_t\}_{t=0,1,\cdots }$ under a history-dependent policy $\pi$ is a controlled Markov chain, with the transition probability:
\begin{align}\label{eqn:trans_prob}
&\Pr[\bs{A}'|\bs{A},s]\\
&=\Pr[\bs{A}_{t+1}=\bs{A}'|\bs{A}_t=\bs{A},s_t=s]\nonumber\\
&=\begin{cases}
1-\lambda,  &~\text{if}~\bs{A}' = (A_d^0,A_r^0) ~\text{and~} s=0,\\
\lambda, &~\text{if}~\bs{A}' = (A_d^1,A_r^0)~\text{and~} s=0,\\
 (1-\lambda)p,&~\text{if}~\bs{A}' =  (A_d^0,A_r^1)~\text{and~} s=1,\\
 (1-\lambda)(1-p),&~\text{if}~\bs{A}' = (A_d^0,A_r^0)~\text{and~} s=1,\\
 \lambda p,&~\text{if}~\bs{A}' = (A_d^1,A_r^1)~\text{and~} s=1,\\
 \lambda(1-p),&~\text{if}~\bs{A}' = (A_d^1,A_r^0)~\text{and~} s=1,\\
0, &~\text{otherwise.}\nonumber
\end{cases}
\end{align}
$A_d^0$ and $A_d^1$ are for the cases that  a new status update arrives and no status update arrives, respectively. $A_r^0$ is for the case that either the device stays idle or the transmission fails, and $A_r^1$ is for the case that the transmission succeeds. From  \eqref{eqn:aoi_device} and \eqref{eqn:aoi_receiver}, we have  $A_d^0= \min\{A_d+1,\hat{A}_d\}$, $A_d^1=1$, $A_r^0=\min\{A_r+1,\hat{A}_r\}$, and $A_r^1=\min\{A_d+1,\hat{A}_r\}$.

For a given history-dependent policy $\pi$, an initial system AoI state $\bs{A}$, and a discount factor $\gamma\in(0,1)$, the infinite horizon expected total discounted AoI at the receiver and the infinite horizon expected total discounted energy cost are, respectively, given by:
\begin{align}
   &\bar{A}_{r,\pi}^\gamma(\bs{A})\triangleq\mathbb{E}\bigg[\limsup_{T\to\infty}  \sum_{t=0}^T \gamma^t A_{r,t}|\bs{A}_0=\bs{A},\pi\bigg],\label{eqn:gamma_aoi}\\   
   &\bar{C}_{\pi}^\gamma(\bs{A})\triangleq\mathbb{E}\bigg[\limsup_{T\to\infty}  \sum_{t=0}^T \gamma^t s_tC|\bs{A}_0=\bs{A},\pi\bigg], \label{eqn:gamma_cost}
\end{align}
where the expectation is taken under the measure induced by policy $\pi$. 
By using the discounted AoI and energy cost, we weight the immediate cost more heavily than expected future costs. 
We use CVaR to capture the  expected tail
loss of the infinite horizon total discounted AoI at the receiver, given by:
\begin{align}
   &\rho_{\pi}^\gamma(\bs{A})\triangleq \text{CVaR}_{\alpha}\bigg(\limsup_{T\to\infty} \sum_{t=0}^T \gamma^t A_{r,t}|\bs{A}_0=\bs{A},\pi\bigg),\label{eqn:gamma_cvar}
\end{align}
where $\alpha\in(0,1]$ is the risk level. 
Note that, $\gamma\in(0,1)$ ensures that $\bar{A}_{r,\pi}^\gamma(\bs{A})$, $\bar{C}_{\pi}^\gamma(\bs{A})$, and $\rho_{\pi}^\gamma(\bs{A})$ are  upper-bounded.

Our goal is to find the optimal history-dependent policy that jointly minimizes the infinite horizon expected total discounted AoI at the receiver, the infinite horizon expected total discounted energy cost, and the CVaR of the infinite horizon total discounted AoI at the receiver. By adopting the weighted-sum method, which a widely used method for multi-objective optimization problem \cite{deb2014multi}, we formulate the following problem:
\begin{align}
\min_{\pi\in\Pi_H}\bar{A}_{r,\pi}^\gamma(\bs{A}) + \eta \rho_{\pi}^\gamma(\bs{A}) + \nu \bar{C}_{\pi}^\gamma(\bs{A}),\label{eqn:risk-mdp}
\end{align}
where $\bs{A}$ is a given initial system AoI state, and $\eta,\nu\geq 0$ are the weighing factors on the CVaR of the AoI and the energy cost. $\eta$ and $\nu$ can be regarded as the penalty factors, mimicking the soft constraints on the CVaR of the AoI and the energy cost. Thus,  we can think $\eta$ and $\nu$ as the corresponding Lagrange multipliers.

We refer to problem \eqref{eqn:risk-mdp} as an infinite horizon discounted risk-aware MDP. Note that, for standard MDPs with expected cost objectives (e.g., \cite{bertsekas4}), it is generally sufficient to focus on the optimization over deterministic stationary Markovian policies without loss of optimality. However, for the considered risk-aware MDP in \eqref{eqn:risk-mdp}, the more general class of history-dependent (non-stationary) policies could be required. 
This is because the CVaR measure is time-inconsistent, which intuitively implies that a policy that is optimal at the current stage is not necessarily optimal in subsequent stages\cite{doi:10.1287/moor.2015.0747}.  
 Such time-inconsistency could further couple risk preferences over time, and, thus prevents us from directly applying dynamic programming to decompose the problem in stages\cite{doi:10.1287/moor.2014.0689}.


\section{Optimal Risk-Aware AoI Solution}
In general, computing the optimal history-dependent updating policy $\pi\in\Pi_H$ for the risk-aware MDP in \eqref{eqn:risk-mdp} is practically intractable due to the substantial requirements in terms of memory and computation time.
 Inspired from \cite{riskphd,doi:10.1287/moor.2015.0747},  we show that the risk-aware MDP in \eqref{eqn:risk-mdp} can be reduced to a standard MDP with an augmented system state space, by exploiting the properties of dual representation and temporal decomposition of coherent risk measures. In particular, the optimal history-dependent policy for \eqref{eqn:risk-mdp} depends on the history only through the augmented system states and can be constructed with the optimal stationary policy for the augmented MDP.

\subsection{Reduction of a Risk-Aware MDP to an Augmented MDP} 
According to \cite[Equation (6.69)]{doi:10.1137/1.9780898718751}, we know that, for any $\alpha,\beta\in(0,1]$, the coherent risk measure $\rho(Z) = (1-\beta) \mathbb{E}[Z] + \beta \text{CVaR}_{\alpha}(Z)$ has the dual representation in the form of \eqref{eqn:dual_rep_risk}, where the risk envelop is
$\Xi = \{\xi:\xi(\omega)\in[1-\beta,1+\beta(1/\alpha -1)], \forall \omega\in\Omega,~\text{and}~\sum_{\omega\in\Omega}\xi(\omega)P(\omega)=1\}.$
Then, we can transform $\bar{A}_{r,\pi}^\gamma(\bs{A}) + \eta \rho_{\pi}^\gamma(\bs{A})$ in the objective function of \eqref{eqn:risk-mdp} to a coherent risk measure:
\begin{align}\label{eqn:obj in coherent}
&\bar{A}_{r,\pi}^\gamma(\bs{A}) + \eta \rho_{\pi}^\gamma(\bs{A})\nonumber\\
&=(1+\eta) \left((1-\frac{\eta}{1+\eta})\bar{A}_{r,\pi}^\gamma(\bs{A}) + \frac{\eta}{1+\eta}\rho_{\pi}^\gamma(\bs{A})\right)\nonumber\\
&\triangleq (1+\eta) \rho_{\delta,\phi}\bigg(\limsup_{T\to\infty} \sum_{t=0}^T \gamma^t A_{r,t}|\bs{A}_0=\bs{A},\pi\bigg),
\end{align}
where $\rho_{\delta,\phi}(\cdot)$ is a coherent risk measure with a risk envelop: $\Xi(\delta,\phi,P) = \{\xi:\xi(\omega)\in[\delta,1/\phi], \forall \omega\in\Omega~\text{and}~\sum_{\omega\in\Omega}\xi(\omega)P(\omega)=1\}$, 
$\delta=\frac{1}{1+\eta}\in(0,1]$ and $\phi = \frac{\alpha(1+\eta)}{\alpha+\eta}\in(0,1]$.
$(\delta,\phi)$ are the risk levels of $\rho_{\delta,\phi}(\cdot)$.

Now, we present the key temporal decomposition property of the coherent risk measure. 
First, for each $k=0,1,\cdots$, we define $\bar{\pi}_k=(\bar{\mu}_t)_{t=k,k+1,\cdots}$ as a $k$-th tail history-dependent policy, where the action $\bar{\mu}_t$ at slot $t\geq k$ is a mapping from $\mathcal{H}_{k,t}$ to the control action space $\mathcal{S}$. Here, $\mathcal{H}_{k,t}$ denotes the set of all histories from slot $k$ to slot $t$, satisfying $\mathcal{H}_{k,t+1} \triangleq\mathcal{H}_{k,t}\times\mathcal{S}\times\mathcal{A}$ for $t\geq k+1$ and $\mathcal{H}_{k,k}\triangleq\bs{A}$. A generic element $\bs{h}_{k,t}$ of $\mathcal{H}_{k,t}$ takes the form $\bs{h}_{k,t}\triangleq (\bs{A}_k,s_k,\cdots,\bs{A}_{t-1},s_{t-1},\bs{A}_t)$. From Definition~\ref{definition:history-dependent policy}, we know that $\bar{\pi}_0=\pi$. 
Then, we have the following temporal decomposition of  $\rho_{\delta,\phi}(\cdot)$,  based on Theorem 2.6.1 in \cite{riskphd}.
%
%
\begin{lemma}\label{lemma:decompostion}
Given slot $k$, system AoI state $\bs{A}_k\in\mathcal{A}$,  control action $s_k\in\mathcal{S}$, and risk levels $(\delta_k,\phi_k)=(\delta,\phi)\in(0,1]^2$, for any $(k+1)$-th tail history-dependent policy $\bar{\pi}_{k+1}$, we have the following temporal decomposition property of the conditional coherent risk measure of $\rho_{\delta,\phi}(\cdot)$:
\begin{align*}
&\rho_{\delta,\phi}(Z_{k+1}|\bs{A}_k,s_k,\bar{\pi}_{k+1}) = \max_{\xi\in\Xi(\delta,\phi,\Pr(\cdot|\bs{A}_k,s_k))} \nonumber\\
&\mathbb{E} \left[\xi(\bs{A}_{k+1})\rho_{\delta/\xi(\bs{A}_{k+1}),\phi\xi(\bs{A}_{k+1})}(Z_{k+1}|\bs{A}_{k+1},\bar{\pi}_{k+1})|\bs{A}_k,s_k\right],
\end{align*}
where $Z_{k+1}\triangleq\limsup_{T\to\infty} \sum_{t=0}^T \gamma^t A_{r,t+k+1}$ denotes the (random) total discounted AoI at the receiver from time $k+1$ such that the system AoI state evolves under policy $\bar{\pi}_{k+1}$ via the AoI dynamics in \eqref{eqn:aoi_device} and \eqref{eqn:aoi_receiver} conditioned on $(\bs{A}_k,s_k)$, and the expectation is taken with respect to the probability distribution of $\bs{A}_{k+1}$ conditioned on $(\bs{A}_k,s_k)$.
\end{lemma}

Note that, the difference between Lemma~\ref{lemma:decompostion} and Theorem 2.6.1 in \cite{riskphd} is that we remove the dependency on the history prior to time $k$. This is because $\bs{A}_k$, $s_k$, and $(\delta_k,\phi_k)$ are given, $Z_{k+1}$ is conditioned on $\bs{A}_{k+1}$, and the system AoI state is Markov. 
Based on the temporal decomposition of the coherent risk measure in Lemma~\ref{lemma:decompostion}, by following the state space augmentation approach in \cite[Chapter 2]{riskphd}, we augment the system AoI state space $\mathcal{A}$ to include additional two dimensional state space $\mathcal{X}\times\mathcal{Y}=(0,1]^2$, which correspond to the two risk levels $(\delta,\phi)$. 
We refer to as $\mathcal{A}\times\mathcal{X}\times{Y}$ as the augmented system state space.
The dynamics of the augmented system state $(\bs{A},x,y)\in\mathcal{A}\times\mathcal{X}\times{Y}$ are as follows: The system AoI states $\{\bs{A}_t\}_{t=0,1,\cdots}$ still evolve as per the AoI dynamics in \eqref{eqn:aoi_device} and \eqref{eqn:aoi_receiver} as well as the transition probability in \eqref{eqn:trans_prob}, and the evolution does not depend on the risk levels. The risk levels $\{x_t,y_t\}_{t=0,1,\cdots}$ evolve deterministically according to $x_{t+1} = x_t/\xi^*(\bs{A}_t,x_t,y_t,s_t)$ and  $y_{t+1} = y_t \xi^*(\bs{A}_t,x_t,y_t,s_t)$, where $\xi^*(\cdot)$ is a known deterministic function that will be specified in \eqref{eqn:xi}. 
Now, we introduce a new class of policies with the augmented system state space. 
\begin{definition}\label{definition:augmented policy}
An \emph{augmented stationary updating control policy} $\tilde{\pi}$ is a sequence of decision rules for each time slot, i.e., $\tilde{\pi}=(\tilde{\mu},\tilde{\mu},\cdots)$, where $\tilde{\mu}$ is a mapping from the augmented system state space $\mathcal{A}\times\mathcal{X}\times{Y}$ to the control action space $\mathcal{S}$, i.e., $s=\tilde{\mu}(\bs{A},x,y)$. Let $\tilde{\Pi}_S$ be the set of all augmented stationary policies $\tilde{\pi}$. 
\end{definition}

Given an augmented stationary policy $\tilde{\pi}$,  an initial augmented system state $(\bs{A},x,y)$, and a discounted factor $\gamma$, we define $\bar{A}_{r,\tilde{\pi}}^\gamma(\bs{A},x,y)$, $\bar{C}_{\tilde{\pi}}^\gamma(\bs{A},x,y)$, and $\rho_{\tilde{\pi}}^\gamma(\bs{A},x,y)$, in the same manner as in \eqref{eqn:gamma_aoi}-\eqref{eqn:gamma_cvar}, respectively, and formulate the corresponding augmented MDP as follows:
\begin{align}
V^*(\bs{A},x,y)\triangleq\min_{\tilde{\pi}\in\tilde{\Pi}_S}V_{\tilde{\pi}}(\bs{A},x,y),\label{eqn:augemented-mdp}
\end{align}
where $V_{\tilde{\pi}}(\bs{A},x,y)\triangleq\bar{A}_{r,\tilde{\pi}}^\gamma(\bs{A},x,y) + \eta \rho_{\tilde{\pi}}^\gamma(\bs{A},x,y) 
+ \nu \bar{C}_{\tilde{\pi}}^\gamma(\bs{A},x,y).$
Next, we  show that the optimal history-dependent updating control policy $\pi\in\Pi_H$ in Definition~\ref{definition:history-dependent policy} for the risk-aware MDP in \eqref{eqn:risk-mdp} can be constructed by obtaining the optimal augmented stationary updating control policy $\tilde{\pi}\in\tilde{\Pi}_S$ in Definition~\ref{definition:augmented policy} for the  augmented MDP in \eqref{eqn:augemented-mdp}.

\subsection{Optimality Equations}
According to \eqref{eqn:obj in coherent} and Lemma~\ref{lemma:decompostion}, for any function $V:\mathcal{A}\times\mathcal{X}\times{Y}\rightarrow\mathbb{R}$, we define the following risk-aware Bellman operator $T: \mathcal{A}\times\mathcal{X}\times{Y}\rightarrow\mathcal{A}\times\mathcal{X}\times{Y}$ on $V$ as follows:
\begin{align}\label{eqn:bellman-operator}
T[V](\bs{A},&x,y) = \min_{s\in\mathcal{S}} \bigg[(1+\eta)A_r + \nu sC \nonumber\\
&\hspace{10mm}+\gamma\max_{\xi\in\Xi(x,y,\Pr(\cdot|\bs{A},s))}\sum_{\bs{A}'\in\mathcal{A}}\Big(\xi(\bs{A}')\nonumber\\
&\hspace{3mm}\times V(\bs{A}',x/\xi(\bs{A}'),y\xi(\bs{A}'))\Pr[\bs{A}'|\bs{A},s]\Big)
\bigg].
\end{align}
Here, from \eqref{eqn:bellman-operator}, we introduce $\xi^*(\cdot)$ as follows:
\begin{align}\label{eqn:xi}
&\xi^*(\bs{A},x,y,s)\triangleq  \arg\max_{\xi\in\Xi(x,y,\Pr(\cdot|\bs{A},s))}\sum_{\bs{A}'\in\mathcal{A}}\Big(\xi(\bs{A}')\nonumber\\
&\hspace{1mm}\times V(\bs{A}',x/\xi(\bs{A}'),y\xi(\bs{A}'))\Pr[\bs{A}'|\bs{A},s]\Big), \forall \bs{A},x,y.
\end{align}

We denote $T^k$ by the composition of the mapping $T$ with itself $k$ times, i.e., $T^k[V](\bs{A},x,y) \triangleq T[T^{k-1}[V]](\bs{A},x,y)$. 
According the definition of the risk envelop of the coherent risk measure $\rho_{\delta,\phi}$, we can show that the  risk-aware Bellman operator $T[V]$ has the monotonicity and constant shift properties\cite[Chapter 1.1.2]{bertsekas4}.\footnote{All  proofs are omitted due to space limitations.}
Given the definition of $T[V]$ in \eqref{eqn:bellman-operator}, we next provide the expression of $T^k[V]$ for $k=1,2,\cdots$.
\begin{lemma}\label{lemma:Tk}
For any $(\bs{A},x,y)\in\mathcal{A}\times\mathcal{X}\times{Y}$ and any function $V:\mathcal{A}\times\mathcal{X}\times{Y}\rightarrow\mathbb{R}$,  we have 
\begin{align}
&T^k[V](\bs{A},x,y) \nonumber\\
&= \min_{\tilde{\pi}\in\tilde{\Pi}_S} \rho_{x,y}\bigg((1+\eta)\sum_{t=0}^{k-1}\gamma^t A_{r,t} + \gamma^k V(\bs{A},x,y| \bs{A}_0=\bs{A},\tilde{\pi}  \bigg)\nonumber\\
&\hspace{6mm}+ \nu \mathbb{E}\bigg[\sum_{t=0}^{k-1} \gamma^t s_tC|\bs{A}_0=\bs{A},\tilde{\pi}\bigg],~\forall k=1,2,\cdots,
\end{align}
where the action $s_t$ is induced by $\tilde{\pi}(\bs{A}_t,x_t,y_t)$.
\end{lemma}

Based on Lemma~\ref{lemma:Tk}, we can obtain the optimal augmented stationary updating control policy $\tilde{\pi}^*$:
\begin{theorem}\label{theorem:find-optiaml-augmented}
For any given $(\bs{A},x,y)\in\mathcal{A}\times\mathcal{X}\times{Y}$, the optimal function $V^*(\cdot)$ in \eqref{eqn:augemented-mdp} satisfies that:
\begin{align}\label{eqn:augmented_bellman}
V^*(\bs{A},x,y)= T[V^*](\bs{A},x,y).
\end{align}
Moreover, $V^*(\cdot)$ is the unique solution to \eqref{eqn:augmented_bellman} within the class of bounded functions.
\end{theorem}

\begin{IEEEproof}[Proof Sketch]
We first show that, for any bounded functions $V:\mathcal{A}\times\mathcal{X}\times{Y}\rightarrow\mathbb{R}$,
\begin{align}\label{eqn:DP-convergence}
V^*(\bs{A},x,y) = \lim_{N\to\infty} T^N[V](\bs{A},x,y),
\end{align}
holds for all $\bs{A},x,y$.
To prove \eqref{eqn:DP-convergence}, we break $V_{\tilde{\pi}}(\bs{A},x,y)$ into the portions incurred over the first $N$ stage and over the remaining stages.
Then, by using the monotonocity and the subadditivity of the coherent risk measure  $\rho_{x,y}(\cdot)$,  the fact $\rho_{x,y}(c)=c$ for a constant $c$, the upper-limit $\hat{A}_r$ of the AoI at the receiver, and Lemma~\ref{lemma:Tk}, we can obtain
\begin{align}\label{eqn:V-TnV}
&|V_{\tilde{\pi}}(\bs{A},x,y)-T^N[V](\bs{A},x,y)|\nonumber\\
&\leq \frac{\gamma^N}{1-\gamma}\big((1+\eta)\hat{A}_r+\nu C + \max_{\bs{A},x,y} |V(\bs{A},x,y)|\big),
\end{align}
based on which, we can prove \eqref{eqn:DP-convergence}.

Next, considering a zero function $V_0(\cdot)$ such that $V_0(\bs{A},x,y)=0$ for all $\bs{A},x,y$, and by the monotonicity and constant shift properties of $T[V]$,
we can show that $V^*= T[V^*]$.

Finally, the uniqueness of the solution to \eqref{eqn:augmented_bellman} can be proved by the monotonicity property of $T[V]$  and \eqref{eqn:DP-convergence}. 
\end{IEEEproof}
Now, we have the  optimal augmented stationary policy $\tilde{\pi}^*$ to the augmented MDP in \eqref{eqn:augemented-mdp}, given by:
\begin{align}\label{eqn:optimal-stationary-policy}
\tilde{\pi}^*(\bs{A},&x,y)= \arg\min_{s\in\mathcal{S}} \bigg[(1+\eta)A_r + \nu sC \nonumber\\
&\hspace{10mm}+\gamma\max_{\xi\in\Xi(x,y,\Pr(\cdot|\bs{A},s))}\sum_{\bs{A}'\in\mathcal{A}}\Big(\xi(\bs{A}')\nonumber\\
&\hspace{3mm}\times V^*(\bs{A}',x/\xi(\bs{A}'),y\xi(\bs{A}'))\Pr[\bs{A}'|\bs{A},s]\Big)
\bigg].
\end{align}
We now show that $\tilde{\pi}^*$ can be used to construct the optimal history-dependent policy $\pi$ for the risk-aware MDP in \eqref{eqn:risk-mdp}.
 
\begin{theorem}\label{theorem:find-optiaml-history-dependent}
For any $\bs{A}\in\mathcal{A}$, $x=\frac{1}{1+\eta}$, and $y=\frac{\alpha (1+\eta)}{\alpha + \eta}$, the optimal function $V^*(\cdot)$ in \eqref{eqn:augemented-mdp} equals to the optimal solution of the risk-aware MDP in \eqref{eqn:risk-mdp}, i.e.,
\begin{align} \label{eqn:equivalent}
V^*(\bs{A},x,y)=\min_{\pi\in\Pi_H}\bar{A}_{r,\pi}^\gamma(\bs{A}) + \eta \rho_{\pi}^\gamma(\bs{A}) + \nu \bar{C}_{\pi}^\gamma(\bs{A}).
\end{align}
Moreover, the optimal history-dependent policy $\pi^*=(\mu^*_0,\mu^*_1,\cdots)$ of \eqref{eqn:risk-mdp}  is given by:
\begin{align}
\mu^*_t(\bs{h}_t) = \tilde{\pi}^*(\bs{A}_t,x_t,y_t),
\end{align}
with the initial system AoI state $\bs{A}_0=\bs{A}$ and risk levels $(x_0,y_0)=(x,y)$. 
Here, the dynamics of $\bs{A}_t$ are given by \eqref{eqn:aoi_device} and \eqref{eqn:aoi_receiver}, and the dynamics of $(x_t,y_t)$ are given by:
\begin{align}
&x_{t+1} = x_t/\xi^*(\bs{A}_t,x_t,y_t,\tilde{\pi}^*(\bs{A}_t,x_t,y_t)),\label{eqn:dynamics_x}\\
&y_{t+1} = y_t\xi^*(\bs{A}_t,x_t,y_t,\tilde{\pi}^*(\bs{A}_t,x_t,y_t)),\label{eqn:dynamics_y}
\end{align}
where $\xi^*(\cdot)$ is given by \eqref{eqn:xi}.
\end{theorem}
\begin{IEEEproof}[Proof Sketch]
First, we show that the optimal solution of the risk-aware MDP in \eqref{eqn:risk-mdp} is also a solution to \eqref{eqn:V-TnV}, by exploiting the $k$-th tail history dependent policy and the temporal decomposition of $\rho_{x,y}(\cdot)$. Then, by the uniqueness of the solution to  \eqref{eqn:V-TnV}, we can immediately have \eqref{eqn:equivalent}. Then, we show that, the history-dependent policy constructed with the associated augmented stationary policy, is optimal, by using similar approaches as in Proposition 1.2.5 in \cite{bertsekas4}.
\end{IEEEproof}

From Theorems~\ref{theorem:find-optiaml-augmented} and~\ref{theorem:find-optiaml-history-dependent}, we observe that, although the original risk-aware MDP in \eqref{eqn:risk-mdp} is defined over the intractable space of history-dependent updating policies, we only need to focus on finding the optimal augmented stationary policy defined in Definition~\ref{definition:augmented policy}, which depends on the original system AoI state $\bs{A}$ and two additional risk levels $(x,y)$.  
Moreover, from the dynamics of the two risk levels $(x_t,y_t)$ in \eqref{eqn:dynamics_x} and \eqref{eqn:dynamics_y}, it can be seen  that the values of $(x_t,y_t)$ contain the historical information that is necessary to make the optimal decision, and thus can be seen as a certain kind of sufficient statistics.
Furthermore, the optimal history-dependent updating control policy $\pi^*\in\Pi_H$ can be derived, by first obtaining the optimal augmented stationary $\tilde{\pi}^*\in\tilde{\Pi}_S$ in \eqref{eqn:optimal-stationary-policy} and then using the construction procedure in Theorem~\ref{theorem:find-optiaml-history-dependent}.
Here, to derive derive $\tilde{\pi}^*$, we can apply the value iteration algorithm\cite{bertsekas4} to obtain $V^*(\cdot)$. Let $V_k$ be the value function at iteration $k$ which is updated  according to $V_{k}(\bs{A},x,y)=T[V_{k-1}](\bs{A},x,y)$. 
By \eqref{eqn:DP-convergence},  under any  initialization of a bounded $V_0(\cdot)$, the generated sequence $\{V_{k}(\bs{A},x,y)\}$ converges to $V^*(\bs{A},x,y)$, i.e., $V^*(\bs{A},x,y) = \lim_{k\to\infty} V_{k}(\bs{A},x,y)$.

In a nutshell, we have proposed a novel approach that explicitly accounts for rare events with very large AoI in a IoT status updating system and developed a dynamic programming based solution to obtain the optimal history-dependent updating policy.
\begin{remark}
The proposed solution framework is significant, as it can be applied to design \emph{optimal solutions} for risk-aware AoI minimization in other IoT scenarios, in which the optimal policy should be history-dependent, and, thus, is generally intractable.
Moreover, the kernel of the proposed solution is dynamic programming, which further allows for the design of more efficient algorithms by levering advanced machine learning in real-time IoT monitoring systems.
\end{remark}
\section{Conclusion}
In this paper, we have studied the optimal process update policy that minimizes the AoI at the receiver, the CVaR of the AoI at the receiver, and the energy cost. 
We have formulated this stochastic optimization problem as an infinite horizon discounted risk-aware MDP.
To obtain the optimal history-dependent policy of the risk-aware MDP, we first reduce it to a standard MDP with an augmented system state space consisting of the original system AoI state space and the state space of two additional risk levels.
For the augmented MDP, we have shown that the optimal stationary policy can be derived through dynamic programming based on a risk-aware Bellman operator.
Then, we have shown that the optimal history-dependent policy of the risk-aware MDP depends on the history only through the augmented system states and can be constructed, by first obtaining the optimal stationary policy of the augmented MDP and then using a special construction procedure.
The proposed solution is shown to be computationally tractable and can be applied in real-time IoT monitoring systems to minimize the AoI.




\bibliographystyle{IEEEtran}
\bibliography{IEEEabrv,ICC20}

\end{document}